\documentclass[aps,prl,twocolumn,showpacs,superscriptaddress]{revtex4}
\usepackage{multirow}
\usepackage{graphicx}
\usepackage{dcolumn}
\usepackage{bm}

\begin{document}


\title{Activated O$_2$ dissociation and formation of oxide islands on the Be(0001) surface: Another atomistic model for metal oxidation}
\author{Yu Yang}
\affiliation{LCP, Institute of Applied Physics and Computational
Mathematics, P.O. Box 8009, Beijing 100088, People's Republic of
China}
\author{Ping Zhang}
\thanks{Corresponding author. zhang\_ping@iapcm.ac.cn}
\affiliation{LCP, Institute of Applied Physics and Computational
Mathematics, P.O. Box 8009, Beijing 100088, People's Republic of
China} \affiliation{Center for Applied Physics and Technology,
Peking University, Beijing 100871, People's Republic of China}


\begin{abstract}
By simulating the dissociation of O$_2$ molecules on the Be(0001)
surface using the first-principles molecular dynamics approach, we
propose a new atomistic model for the surface oxidation of $sp$
metals. In our model, only the dissociation of the first oxygen
molecule needs to overcome an energy barrier, while the subsequent
oxygen molecules dissociate barrierlessly around the adsorption
area. Consequently, oxide islands form on the metal surface, and
grow up in a lateral way. We also discover that the firstly
dissociated oxygen atoms are not so mobile on the Be(0001) surface,
as on the Al(111) surface. Our atomistic model enlarges the
knowledge on metal surface oxidations by perfectly explaining the
initial stage during the surface oxidation of Be, and might be
applicable to some other $sp$ metal surfaces.

\end{abstract}

\maketitle

Most metal surfaces are thermodynamically allowed to react with
oxygen molecules in the atmosphere, to form thin oxide films
\cite{Lawless1974,Ciacchi04}. This phenomenon is very important for
many technological applications, because they are directly relevant
to surface corrosion and the formed metal oxide films have been
widely used as catalysts, sensors, dielectrics, and corrosion
inhibitors \cite{Kung1989}. In addition, studying the oxidation
reactions from the atomic view is also of great scientific
importance \cite{Busnengo00}. However, the detailed atomistic models
for the surface oxidation of metals are not yet mature
\cite{Ciacchi04}. A ``Hot-Atom'' O$_2$ dissociation and oxide
nucleation model was recently proposed for the oxidation of the
Al(111) surface, in which the dissociative adsorption of O$_2$
molecules occurs via a ``Hot-Atom'' mechanism and O atoms are
spontaneously incorporated underneath the topmost Al surface layer,
initiating the nucleation of the oxide far below the saturation
coverage of one O adlayer \cite{Ciacchi04}. However, the
applicability of this model has not been checked for other metals,
for example, whether the dissociated O atoms on other metal surfaces
are ``hot'' is still beyond the present knowledge. Besides, there is
a contradiction with experimental observations in the atomistic
model. The initial sticking probability of thermal O$_2$ molecules
at Al(111) is measured to be low by many independent experiments
\cite{Brune1993}, suggesting a sizeable energy barrier for the O$_2$
dissociation, which however is not included in the ``Hot-Atom''
mechanism. The lack of an energy barrier is contributed to the fact
that in the adiabatic first-principles calculations the lowest
unoccupied electronic state of oxygen is aligned with the Fermi
level at any distance between the molecule and the surface, allowing
a partial filling of the empty molecular orbital, which finally
drives the dissociation \cite{Ciacchi04}.

Based on this background, in this paper we have carried out
first-principles molecular dynamics (FPMD) simulations for the O$_2$
dissociation on the Be(0001) surface. Chemically, Be is even simpler
than Al because it has only two valence electrons, so it is also an
ideal model system to study the initial stages of metal oxide
formation. Besides this basic point of interest, our present study
is also motivated by the fact that Be has vast technological
applications due to its high melting point and low weight
\cite{Zalkind1997}. During these applications, surface oxidation as
the main kind of corrosion always needs to be prevented.
Experimentally, the surface oxidation of Be is reported to begin by
forming separate oxide islands, and saturate after the islands grow
laterally together forming an oxide layer \cite{Zalkind1997}.
Theoretically, we have calculated the adsorption properties of O$_2$
molecules on the Be(0001) surface \cite{Zhang09}, and identified
both the physisorbed and chemisorbed molecular precursor states.
Remarkably, we have revealed that unlike the Al(111) surface, the
alignment of the lowest unoccupied electronic state of oxygen with
the Fermi level does not happen when it is in close with the
Be(0001) surface \cite{Zhang09}. Therefore, the Be(0001) surface is
in some way a better model system for studying the metal oxidation
using adiabatic first-principles methods.

Our calculations are performed using the spin-polarized version of
the Vienna {\it ab-initio} simulation package \cite{VASP}. The PW91
\cite{PW91} generalized gradient approximation and the
projector-augmented wave potential \cite{PAW} are employed to
describe the exchange-correlation energy and the electron-ion
interaction, respectively. The cutoff energy for the plane wave
expansion is set to 400 eV. The molecular dynamics (MD) simulations
are performed using the Verlet algorithm with a time step of 1 fs
within the micro canonical ensemble. In our present study, the
Be(0001) surface is modeled by a periodically repeated slab of five
Be layers separated by a vacuum region correspondent to six metal
layers. We consider a (4$\times$4) surface unit cell, which includes
16 Be atoms in each atomic layer. The surface Brillouin zone is
sampled by a 3$\times$3 $k$-point distribution using the
Monkhorst-Pack scheme \cite{Monkhorst}. The calculated lattice
constant of bulk Be and the bondlength of isolated O$_{2}$ are 5.03
\AA~ and 1.24 \AA, respectively, in good agreement with the
experimental values of 4.95 \AA~\cite{Wyckoff1965} and 1.21
\AA~\cite{Huber1979}. The O$_{2}$ is placed on one side of the slab,
namely on the top surface, whereas the bottom two layers are fixed.
All other Be layers as well as the oxygen atoms are free to move
during the MD simulations.

We start our simulations with different orientations of an O$_2$
molecule placed over different surface sites (one representative
case is illustrated in Fig. 1(a)). In all initial configurations,
the mass center of the O$_2$ molecule is initially set to be 4 \AA~
away from the metal surface. In the cases of Al(111) and Mg(0001),
the failure of adiabatic FPMD simulations in producing an
activated-type dissociation process has been ascribed to the
unphysical output that charge transfer occurs at any molecule-metal
distance, which has led to speculations that nonadiabatic effects
may play an important role in the oxygen dissociation process at
these metal surfaces with simple $sp$ electrons
\cite{Behler05,Zhang09}. However, this requirement is not always
needed. In our calculations, we find that at enough molecule-metal
distance, the unphysical large-distance charge-transfer effect does
not happen between O$_2$ and the Be(0001) surface. The calculated
spin-split electronic states of O$_2$ in close to the Be(0001)
surface shows no change with respect to the free molecule. The
lowest unoccupied molecular orbital remains empty and the
charge-density difference is zero everywhere. Subsequently, the
molecular bond length and spin of O$_2$ are not influenced at all by
the presence of the Be(0001) surface \cite{Zhang09}. Thus we are
confident in carrying out our present first-principles molecular
dynamics simulations.

\begin{figure}[tbp]
\begin{center}
\includegraphics[width=0.7\linewidth]{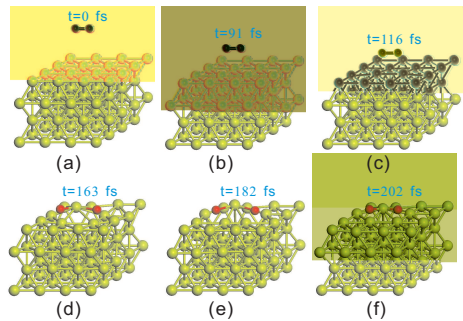}
\end{center}
\caption{(color online). Snapshots from a FPMD simulation of the
dissociative chemisorption of an O$_2$ molecule on the Be(0001)
surface. Only the four outermost Be layers are shown. Red and yellow
balls respectively represent oxygen and Be atoms. (a) Initial
geometry (simulation time $t$=0 fs) with the molecule parallel to
the surface. The distance between the center of mass of the molecule
and the topmost surface layer is 4.00 \AA. (b)-(e) The atomic
geometry of the adsorption system at $t$=91, 116, 163, and 182 fs.
(f) Final configuration in equilibrium after relaxation ($t$=202
fs). The final O-O distance is 2.33 \AA.}
\end{figure}%

For the adsorption of the first O$_2$ molecule, the substrate atoms
are initially at rest. Since we have known from our static
calculations that an energy barrier is needed for the dissociation
of O$_2$ molecule on the clean Be(0001) surface \cite{Zhang09}, we
set two different initial kinetic energies for the O$_2$ molecule,
respectively 0.06 eV and 0.6 eV. From the MD simulations, we find
that the O$_2$ molecule with the initial kinetic energy of 0.06 eV
does not dissociate at all on the Be(0001) surface after 3 ps, but
the one with the initial kinetic energy of 0.6 eV easily dissociate
in 200 fs. These results clearly prove that an energy barrier exists
during the O$_2$ dissociation on the Be(0001) surface, with the
value between 0.06 and 0.6 eV. The structural evolution for the
O$_2$ dissociation with the initial kinetic energy of 0.6 eV is
depicted in Fig. 1. We can see that the surface Be atoms have no
motions until $t$=116 fs, when only electronic interactions happen.
From the time $t$=116 fs to 163 fs, one surface Be atom is pulled
out a little during the departure of the two oxygen atoms. After
$t$=116 fs, the two oxygen atoms steadily adsorb at the two hcp
hollow sites, and the adsorption system begins to vibrate in its
intrinsic frequencies.

\begin{figure}[tbp]
\begin{center}
\includegraphics[width=0.7\linewidth]{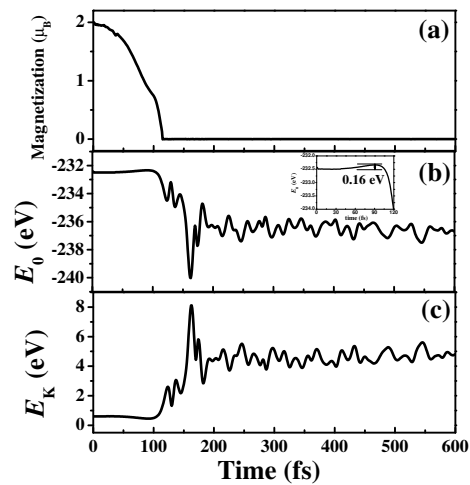}
\end{center}
\caption{(a) The total spin magnetization of the system in a FPMD
simulation of the dissociative chemisorption of an O$_2$ molecule on
the Be(0001) surface. (b) and (c) The total free energy and kinetic
energy of the system in the same FPMD simulation. The inset in (b)
is zoomed out to show the 0.16 eV energy barrier during the O$_2$
dissociation.}
\end{figure}%

The FPMD simulations for the O$_2$ dissociation are subtle. We find
in our testing calculations that using unappropriate mixing
parameters might lead to unphysical change of O$_2$ from the triplet
to singlet state, and resulting wrong trajectories. With the
appropriate mixing parameters, we get the true dissociation process
as shown in Fig. 1 and the natural spin evolution as shown in Fig.
2(a). It is clearly shown that the total spin of the system
gradually decrease into zero as the O$_2$ molecule gets closer to
the Be surface and dissociates. We can also see that the
dissociation process and the electronic evolution go on at the same
time.

At the beginning of the adsorption process, the total free energy of
the adsorption system is $-$232.49 eV. When the O$_2$ molecule gets
to be 2.36 \AA~ from the Be surface, the free energy is enlarged by
0.16 eV [as shown in the inset of Fig. 2(b)], which corresponds to
the energy barrier for the O$_2$ dissociation on the clean Be
surface. The value differences with our static potential energy
surface results \cite{Zhang09} come from that we have chosen a much
larger $4\times4$ supercell, and the motions of surface Be atoms are
considered here. Accompanying with the enlargement of the total
energy, the kinetic energy decreases by 0.16 eV, as shown in Fig.
2(c). After that, the kinetic energy begins to fluctuate, and reach
the equilibrium value of 4.52 eV at $t$=202 fs. At the same time,
the total free energy of the adsorption system becomes $-$236.41 eV,
which is 3.92 eV below the initial value of the adsorption process.

Although the adsorption energy is large, we see no movement of the
dissociated oxygen atoms during simulation process. Instead, the
oxygen atoms steadily adsorb at two neighboring hcp hollow sites and
vibrate. Comparing with the hot oxygen atoms on the Al(111) surface,
which move away very fast after dissociation \cite{Ciacchi04}, the
dissociated oxygen atoms on the Be(0001) surface are not so hot,
without any fast movements. Thus, our finding for the dissociation
mechanism of O$_2$ is very different from the ``Hot-Atom'' picture,
which was theoretically understood mostly through simulating the
O$_2$/Al(111) prototype \cite{Ciacchi04}. The different dissociation
mechanisms may come from different surface electronic structures of
Al and Be. In particular, our result that the dissociated oxygen
atoms have low mobility is very consistent with the experimental
observations that the surface oxidation of Be begins by forming
separate nucleation islands.

\begin{figure}[tbp]
\begin{center}
\includegraphics[width=0.7\linewidth]{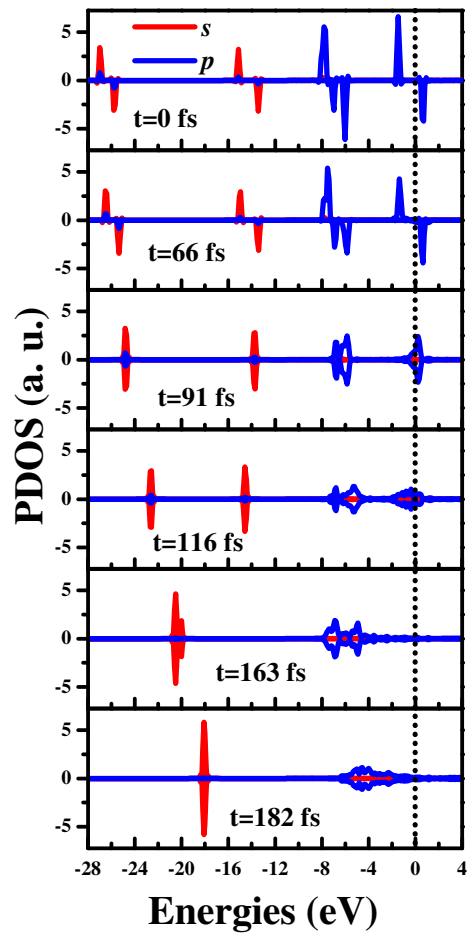}
\end{center}
\caption{(color online). Spin-resolved densities of states around
the two oxygen atoms at different times, in a FPMD simulation of the
dissociative chemisorption of an O$_2$ molecule on the Be(0001)
surface. The $s$ and $p$ electronic states are shown respectively in
red and blue lines.}
\end{figure}%

The evolution of the spin configurations during the adsorption and
dissociation of O$_2$ molecules on metal surfaces is of great
theoretical importance \cite{Behler05,Zhang09}. Since the adiabatic
calculations are suitable enough for the O$_2$/Be(0001) system, we
would like to take a look at the evolutions in spin configurations
of O$_2$ during the dissociative adsorption process. Figure 3 shows
the spin-resolved densities of states for the two oxygen atoms at
different times. We can see that at the very beginning, the O$_2$
molecule is in the triplet state, with the spin splittings of about
2 eV for both bonding and antibonding orbitals. Until $t$=66 fs, no
big changes happen in the electronic structures of O$_2$, except
that the spin splitting decreases about 0.2 eV for all orbitals.
After that, the total spin starts to distribute both around oxygen
and around the neighboring Be atoms. Then at the transition state
($t$=91 fs), the spin splitting decreases to be negligible around
each oxygen atom. We can see that the spin quenching effect from
electronic hybridizations between O$_2$ and the Be(0001) surface
happens really fast. After crossing the transition state, the total
free energy quickly goes down since the two oxygen atoms are
separated and begin to bond with surface Be atoms. At the meantime,
the energy difference between bonding and antibonding orbitals of
O$_2$ reduces and finally to be zero at $t$=182 fs. In total, there
are two stages during the O$_2$ dissociation. At the first stage
(i.e. before the transition state), the electronic interaction
results in spin quenching and makes the two oxygen atoms to separate
from each other. At the second stage, the two oxygen atoms begin to
bond with Be atoms, blurring the distinction in bonding and
antibonding orbitals.

It has been experimentally reported that the surface oxidation of Be
begins by forming separate oxide islands. There might be two
different mechanisms responsible for this observation: i) The
dissociated oxygen atoms are very mobile and tend to cluster with
each other; ii) The subsequent O$_2$ molecules are easier to
dissociate around the oxygen-adsorbed Be surfaces. The first
mechanism has been proposed to explain the oxide nucleation on the
Al(111) surface \cite{Ciacchi04}. Since we have revealed that the
dissociated oxygen atoms are not so mobile on the Be(0001) surface
as on the Al(111) surface, the formation of oxide islands on the
Be(0001) surface should have a different mechanism (mechanism ii).
We then simulate the adsorption of a second O$_2$ molecule on the
Be(0001) surface.

\begin{figure}[tbp]
\begin{center}
\includegraphics[width=0.7\linewidth]{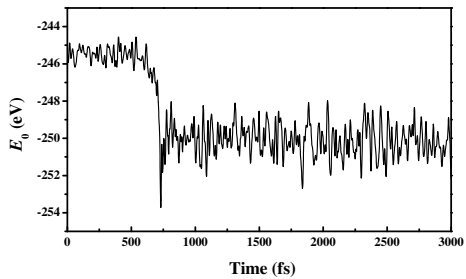}
\end{center}
\caption{The total free energy of the adsorption system in the FPMD
simulation for the dissociative adsorption of the second O$_2$
molecule.}
\end{figure}%

For the adsorption of the second O$_2$ molecule, the substrate Be
and O atoms are no longer at rest. Their initial velocities are
taken from the equilibrium state of the MD simulation for the first
O$_2$ molecule \cite{2ndO2}. At the same time, the second O$_2$
molecule is initially set at rest, i.e., with zero initial kinetic
energy. After the dissociation of the first O$_2$ molecule, the
temperature within the $4\times4$ surface area raises by about 700
K. If the second O$_2$ molecule approaches the same surface area, it
will be activated from the triplet groundstate to the singlet state
under such a high temperature. From our FPMD simulation, we see that
the second O$_2$ molecule dissociates easily in 600 fs, and the
total free energy decrease about 4.76 eV, as shown in Fig. 4.
Therefore, we can see that the dissociation of the second O$_2$
molecule can be motivated by the adsorption energy of the first
dissociated O$_2$ molecule. Since the adsorption energy of the
second O$_2$ molecule will further enlarge the temperature of the
same surface area, we can foretell that a subsequent third O$_2$
molecule will also dissociate barrierlessly. At last, the O$_2$
molecules always dissociate easier around the oxygen covered Be
surfaces to form separate oxide islands. And if more oxygen
molecules are introduced, the separate oxide islands will grow
larger, until finally meet each other laterally. One can see that
our atomistic model can perfectly explain the experimental results
on the surface oxidation of Be.

Looking after the surface oxidation of another $sp$ metal lead (Pb),
we can find very similar characters. Firstly, the initial sticking
coefficient of O$_2$ is observed to be very low
\cite{Thurmer02,Ma07}, indicating that the O$_2$ dissociation is an
activated process. Secondly, the binding energy per oxygen atom on
the Pb(111) surface is 4.75 eV \cite{Sun08}, also implying that
O$_2$ molecules are easier to dissociate around previously adsorbed
oxygen atoms. In fact, the oxidation process of the Pb(111) surface
has already been observed to be autocatalytic \cite{Thurmer02}, in
accordance with our atomistic model that the oxide islands grow
laterally to be oxide films. All these similarities make us believe
that our atomistic model might be also applicable to some other $sp$
metals.

In summary, we have systematically investigated the adsorption and
dissociation of O$_2$ molecules on the Be(0001) surface within
spin-polarized first-principles molecular dynamics simulations. We
find that the previous ``Hot-Atom'' mechanism are not suitable for
the O$_2$/Be(0001) system. The O$_2$ dissociation is found to be an
activated process on the Be(0001) surface, and the dissociated
oxygen atoms are not very mobile. Based on our calculational
results, we propose a new atomistic model for the surface oxidation
of $sp$ metals, in which the dissociation of the first O$_2$
molecule needs to overcome an energy barrier, while the subsequent
O$_2$ molecules dissociate barrierlessly at the same surface area.
In this way, separate oxide islands form on the metal surface, and
grow up together in a lateral way as more oxygen molecules are
introduced.

\begin{acknowledgments}
This work was supported by the NSFC under grants No. 10904004 and
No. 60776063.
\end{acknowledgments}


\begin{thebibliography}{}
\bibitem{2ndO2}
Even if the substrate O and Be atoms are initially at rest, our
calculations find that the secondly introduced O$_2$ molecule can
dissociate barrierlessly.
\end{thebibliography}
\end{document}